# Dual-band Harmonic and Subharmonic Frequency Generation Circuitry for Joint Communication and Localization Applications Under Severe Multipath Environment

Payman Pahlavan, *Student Member, IEEE*, and Najme Ebrahimi, *Member, IEEE*

*Abstract—* **The next generation of ultra-dense connected and automated wireless sensor networks (WSN) requires proximity intelligence for many of its applications, especially for identification and localization. This work presents the first bidirectional circuitry for Internet of Things (IoT) transponder that reciprocally generates harmonics and subharmonics, dual band frequencies. A multi-band or wideband localization system is essential for future intelligent WSN to mitigate the effect of multipath signals in indoor dense environments. The proposed frequency generation circuitry is based on the novel nonlinear ring resonator (NRR) operating based on standing wave resonation. The proposed NRR generates two sustainable oscillation frequencies based on the periodicity of the nonlinear circuit in the ring configuration. Due to the symmetry and reciprocity of the ring layout, the two bidirectional ports can excite the circuit at the two opposite nodes while maintaining the required boundary conditions for oscillation. The sustainable resonance conditions occur by creating zero, short impedance, and pole, infinite impedance, at subharmonic and harmonic excitation ports. The NRR circuit consumes zero DC power and covers two communication frequency plans interchangeably, which makes it a premier technique compared to the conventional ultra-wideband (UWB) localization system and conventional single band nonlinear passive circuitry. The latter is narrowband due to the tunning limitation of the nonlinear varactor while the former is a power hungry approach with complex hardware requirements. The circuit has 10 dB conversion loss with 200 MHz bandwidth in subharmonic generation or divider mode, and 18 dB loss in the harmonic generation or doubler mode. The average output power of the divider with 0-dBm input power is -9 dBm at $f_{in}$ = 4.8 GHz and -16 dBm for doubler at $f_{in}$ = 2.4 GHz, supporting in the order of few meter level link distance for joint communications and localization.**

## I. INTRODUCTION AND MOTIVATION

THE next generation of intelligent wireless world, 5G and beyond, requires radio technologies such as radio frequency identification (RFID) for accurate indoor localization and ranging and multi-node connectivity [1-5]. The network of over one trillion Internet of Things (IoT) devices connected in the future smart wireless world demand centimetre-level accuracy for identification, ranging, and joint localization and communication in many applications including asset tracking, motion and health monitoring [6, 7], robotic feedback control [8], gesture recognition and acquisition in human machine interface [9] and distributed collaborative communications and sensing [10-13], etc.

Accurate localization and ranging becomes increasingly challenging as the number of connected nodes increases, mainly due to the rich scattering indoor environments leading to severe fading, crosstalk, and pulse dispersion. The undesired multipath interference signals as well as the transmitter (TX) to receiver (RX) self-interference leakage, demonstrated in Fig. 1(a), reduce the positioning accuracy dramatically. The available ranging techniques such as received signal strength (RSS) [14-16], time of arrival (TOA) [17,18], or phase of arrival (POA) [19], suffers significantly from these nonidealities, causing phase ambiguity in POA or TOA and poor accuracy and reliability of main signal extraction in RSS method. Therefore, localization algorithms are developed to improve the localization accuracy and adequate range measurement.

In order to mitigate the multipath effect and to distinguish all the individual multipath components (MPCs) as well as estimating the fading points, the wideband localization techniques, such as ultra-wideband (UWB) signals [20], are required. The main drawback of UWB technology is imposing the higher hardware costs and power consumption on the overall system compared to the passive nonlinear based IoT tags [21-45]. Therefore, solutions based on the UWB system are not suitable for the miniaturized and low-power applications. On the other hand, the nonlinear passive tags are also inherently narrowband, and the ranging error significantly depends on the fading points of the channel. In addition to multipath effect, the self-interference leakage issue between the TX and RX of the reader has been addressed by introducing harmonics [21-27] or subharmonic generating circuitries [28-37] as the IoT tag transponder which separate the transmitted and received frequency bands, shown in Fig. 1(b).

In this work, for the first time, a dual band harmonic and subharmonic system is proposed to cover wider bandwidth, two bands, at both uplink and downlink interchangeably, while separating the TX and the RX signals for self-interference cancellation. The proposed protocol estimates both uplink and down link channels symmetrically at two harmonic and subharmonic bands, *f* and *2f*, shown in Fig. 1. (b). In other words, the reader switches the TX signal in two bands, *f* and *2f*, while the proposed novel circuitry generates and send back *2f* (harmonics), and *f*, (subharmonics), respectively. The localization phase error, $\Delta\theta$, defined as $\arctan(Im(H)/Re(H))$, in [17] (*H* is the channel response), is calculated and compared with the conventional single band frequency generation protocol and is illustrated in Fig. 1(c) for multiple numbers of multipath. The proposed dual band system will improve the localization error significantly compared to the single band harmonic generation as the number of multipath increases due the bandwidth enhancement.

There are several circuit architectures to separate uplink and downlink bands on the IoT tag. The down/up conversion of the incoming signal with a mixer and local oscillator (LO) in homodyne or heterodyne circuitries is a power-hungry option as it requires sophisticated phase and frequency synchronization between the reader and the tag [38]. The self-oscillating mixers based on the injection locking phenomena also consume beyond the energy harvesting level of an IoT sensor with a limited communication distance [39-42, 47-49]. Therefore, the passive frequency generation based on the nonlinearity of varactor diodes is reported as a better candidate for future low-power and miniaturized IoT sensors [21-23]. Although the DC power consumption of nonlinear passive circuitry can be optimally zero, they suffer from low conversion loss. The periodic nonlinear harmonic generation circuit has been proposed to enhance the power efficiency of the system through optimizing the number of cascaded stages [24,27]. However, none of these approaches are reciprocal or bidirectional to cover multi- band frequency and concur the multipath effect. In this work, a novel fully passive zero-bias circuity is proposed that generates both harmonics and subharmonic using the single circuitry to cover two bands for accurate localization in multipath environment. The proposed idea is based on standing wave oscillation produced by two resonance conditions at harmonics and subharmonics using non-linear a ring structure. The novel nonlinear ring resonator (NRR) oscillates at two frequency bands by creating zero, short impedance, and pole, infinite impedance, at divider and doubler input excitation ports, respectively. Section II. A and B describe the harmonics and subharmonics signal generations based on the nonlinearity concepts of varactors. The simultaneous and bidirectional harmonic and subharmonic generation by the proposed NRR architecture is discussed in Section II. C and D. Section III presents the hardware implementations and measurement results. Section IV concludes the paper.

## II. PROPOSED DUAL BAND HARMONIC/SUBHARMONIC GENERATION CIRCUITRY BASED ON NONLINEAR RING RESONATOR (NRR)

### A. Frequency Generation Based on Nonlinear Concept

Harmonics or subharmonic generation circuitry has been conventionally based on nonlinear characteristics of passive components such as varactors [43-45]. Nonlinear behavior of the diode capacitance with respect to the applied small signal voltage can generate higher harmonics as shown in Fig 2 (a). On the other hand, pumping the nonlinear diode with a large signal can also introduce negative resistance and initiate oscillation at subharmonic frequency under required resonance condition, Fig 2 (b). In this work, for the first time, we are going to design a circuity based on both nonlinear features to generate harmonic and subharmonic simultaneously.

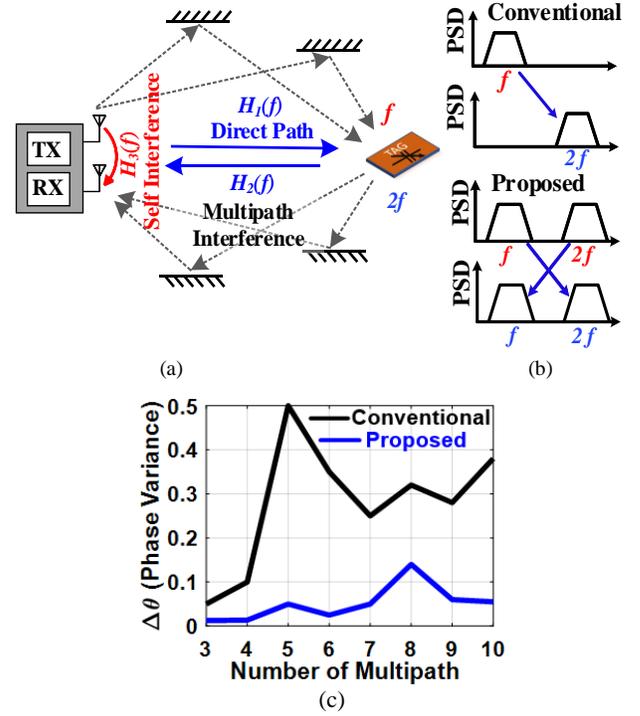

Figure 1. Frequency generation schemes for localization under severe multipath environment. (a) Harmonic generation in single band at the IoT tag, (b) proposed dual band, harmonic and subharmonic generation at IoT tag. (c) Phase variance comparison for proposed scheme under different number of multipath obstacles.

For harmonic generation, nonlinearity of a varactor's capacitance is leveraged to produce harmonics of the input frequency. The capacitance of a p-n junction varactor with a zero-bias voltage capacitance $C_0$, junction voltage $V_j$, and the grading factor $m$ can be defined as [32]

$$C(V) = \frac{C_0}{\left(1+\frac{V_{DC}}{V_j}\right)^m} \quad (1)$$

where $V_{DC}$ is the DC reverse voltage applied to the varactor diode. Since the capacitor current is $i_d = \frac{d(CV)}{dt}$, by applying a sinusoidal small signal $v_s = V_S \sin \omega t$ with $\omega = 2\pi f t$ across the diode, higher harmonics of the input frequency appear in the diode current as

$$I_d = I_{DC} + \omega C_0 V_S \cos \omega t + \omega C_1 V_S^2 \sin 2\omega t + \cdots \quad (2)$$

In order to deliver optimum power to the varactor and suppress unwanted harmonics, the varactor needs two passive filter circuits at both input and output, as shown in Fig. 2(a), to provide sufficient matching and out of band suppression. The conversion loss of the harmonic based multiplier is relevant to the insertion loss of these passive filters as well as the nonlinearity of the varactor.

For subharmonic generation, however, single varactor in frequency dividers introduces capacitance variation with respect to the applied pump voltage to create a negative resistance and provide oscillation condition at the subharmonic frequency. It has been shown in [31] that under a large pump signal injection at incoming frequency $2f$ and with the amplitude of $V_P$ as $v_P(t) = V_P \sin(2\omega t)$, the voltage dependent capacitance $C(V)$ can be modulated at the same injected frequency and create a negative resistance at the subharmonic frequency, $f$, due to derivative relation of $C(V)$ in (1). This equivalent negative impedance, $R_e$, is required for oscillation in a parallel RLC tank integrated in the input/output matching network as shown in Fig. 2 (b). To find out the negative resistive equation, the current and voltage relation at subharmonic should be derived by assuming that the circuit has a capacitor with nonlinear relation as $C(v_P)$ which only varies by the large pump signal $v_P$. The current across the diode, $i_d$, produced by the subharmonic small voltage, $v_s(t) = V_S \sin \omega t$, can be calculated as.

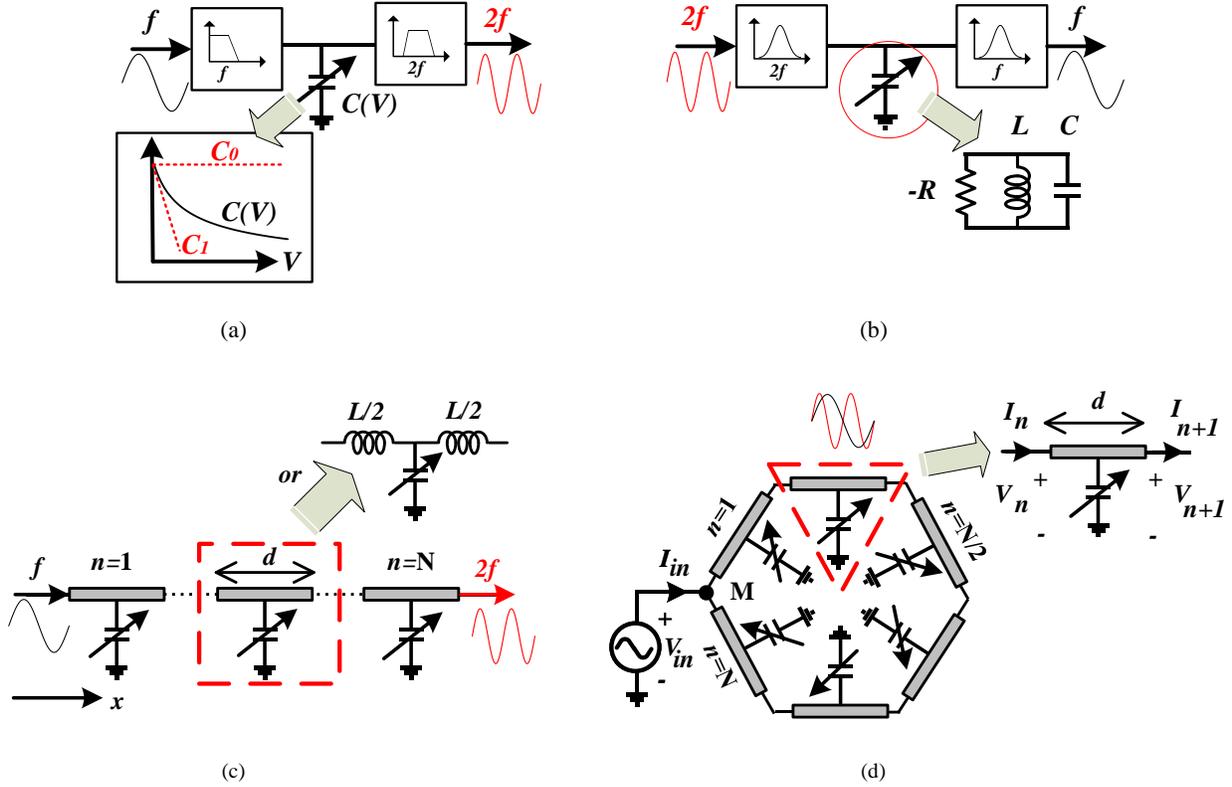

Figure 2. Nonlinear passive circuits working as frequency multiplier and divider. a) Single varactor frequency doubler using voltage variant capacitance of a diode to produce higher harmonics. b) Single varactor frequency divider producing a negative resistance in an LC tank at subharmonic frequency c) Nonlinear transmission line frequency doubler based on periodic LC circuit used to minimize the conversion loss. d) Proposed Nonlinear Ring Resonator (NRR) based on periodic varactor loaded transmission line providing harmonic resonant frequencies as a bidirectional dual band harmonic/subharmonic generator circuitry.

$$i_d = \frac{d}{dt}\{C(v_P) \times v_s(t)\}$$
$$= \frac{d}{dt}\left\{C_0\left(1 + \frac{1}{v_J}V_P \sin 2\omega t\right)^{-m} \times V_S \sin \omega t\right\} \quad (3)$$

By neglecting all other harmonics and sorting $i_d$ as a linear combination of $v_s$ and $dv_s/dt$, the behaviour of the varactor resembles an effective capacitor $C_e$ in parallel with an effective resistor $R_e$ approximated as below (see. Appendix)

$$i_d \approx C_e \frac{d}{dt}V_S \sin \omega t + \frac{1}{R_e}V_S \sin \omega t \quad (4.a)$$

$$C_e = C_0 + \frac{C_2 V_P^2}{2} \quad (4.b)$$

$$R_e = -2/\omega C_1 V_P \quad (4.c)$$

where the varactor capacitance is approximated by first order Taylor expansion with $C_0$ and $C_1$ coefficients. The negative resistance is inversely related to $\omega C_1 V_P$ meaning that increasing either $C_1$, nonlinearity of diode, or $V_P$ can provide a larger conductance for the subharmonic current generation. Furthermore, to deliver efficient pump power, $V_P$, to the varactor and to reduce the insertion loss of the passive network, LC resonator are needed to provide optimum power matching at the two input and output ports, while sustaining the resonant conditions. On the other hand, providing the power matching at the subharmonic frequency will increase the load effect on the LC tank, translated as a parallel positive resistor, leading to a larger threshold power, $P_{th}$ to initiate the subharmonic oscillation, [31-32]. Therefore, there is a trade-off between the minimum input threshold power, $P_{th}$, requirement and the conversion loss of such subharmonic divider based on passive filter using a single varactor.

In order to mitigate the conversion loss and minimum threshold power trade-off, nonlinear transmission line (NLTL) approach has been proposed [26-27]. This approach employs multiple varactors in a periodic ladder configuration, providing additional degree of freedom, number of stages, $n=N$, in Fig. 2(c), for designing multiplier [26-27] or divider [34-35]. The degree of freedom is given through separating the input and output matching filters from the intermediate stage for sustaining simultaneous resonance condition and maximum power transfer. The periodic configuration of this circuitry is based on the travelling wave concept and therefore the waveform is generated through constructive superposition of travelling waves from multiple varactors, $n$, along the propagation directions. In this work, we aim to use such periodic structure to design a circuitry for bidirectional harmonic and subharmonic signal generations for RFID IoT tag. Another main advantage of the periodic capacitive loaded configurations is the miniaturized area due to slow-wave property caused by larger propagation constant, $\beta$ [46].

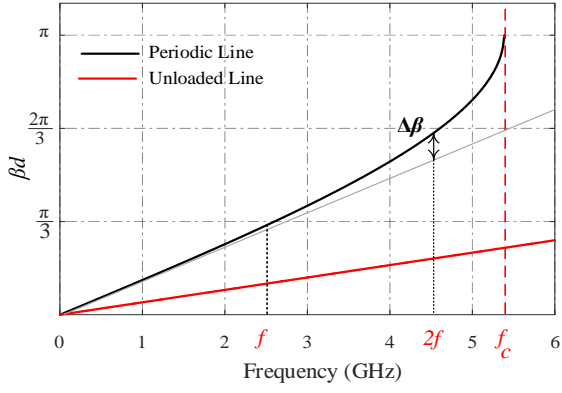

Fig.3. Comparison between propagation delay, $\beta d$, of a unit-cell in a NLTL and a microstrip transmission line with the same length. The dispersion effect on the propagation constant is shown by $\Delta\beta$ while $f_c$ is the cutoff frequency of the periodic line.

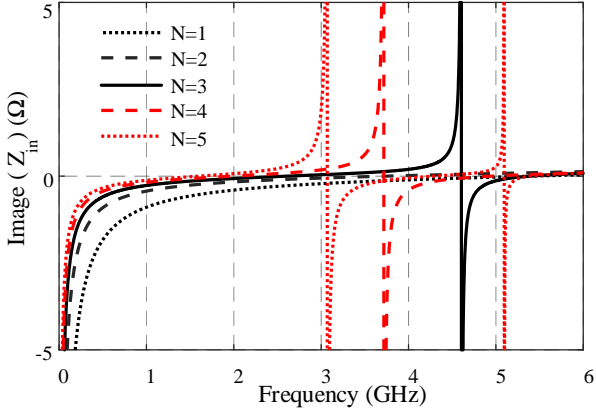

Fig 4. Resonance behavior of the input impedance of the proposed nonlinear ring resonator (NRR) in Fig. 2.(d) based on the total number of unit-cells, $N$.

To analyse the effect of the number of stages, $n=N$, in the design of harmonic generator or subharmonic generator, one can study a simple doubler in NLTL configuration. For constructive harmonic generation through a periodic configuration, each unit cell contains shunt varactors between two transmission line with width of $d/2$, or equivalently with lumped inductors with inductance of $L/2$, as shown in Fig. 2 (c). Each of these varactors in the transmission line introduces harmonics of the excitation signal propagating throughout the line. The amplitudes of each harmonic at every node can be calculated by assuming that the $i$-th, ($i= 1, 2, 3, ...$), harmonic is also propagating along the NLTL in $x$ direction with propagation and attenuation constants $\beta_i$ and $\alpha_i$. In addition, by assuming the subharmonic signal tone at $f$ as $V_s(x) = V_s e^{-(\alpha_1 + j\beta_1)x}$, the second harmonic voltage amplitude, $V_p$, at each unit-cell node can be calculated as a function of $K_{C(v)}$, diode nonlinearity explained in [44], the line dispersion, $\Delta\beta = (\beta_2 - 2\beta_1)$, and the number of stages, $n=N$, as

$$V_p(n) \cong K_{C(v)} V_s^2 \frac{j\beta_2 nd}{B_2 - 2B_1} \sin\frac{(\beta_2 - 2\beta_1)nd}{2} e^{-(\alpha_2 + j\beta_2)nd} \quad (5)$$

where $\Delta\beta = (\beta_2 - 2\beta_1)$ is a coefficient to consider the dispersion effect on the propagation constant, $\beta$, in a dispersive line, which indicates doubling the frequency does not proportionally result in $2\beta$ due to the dispersion effect shown in Fig. 3. In addition, the second harmonic signal term has a decaying sinusoidal behaviour as a function of $n$, shown as $\sin\frac{\Delta\beta nd}{2} e^{-\alpha_2 n}$ in (5). While the decaying term limits the maximum number of unit-cell to achieve the optimum conversion loss, the sinusoid term can be maximized for any integer value of $n$ satisfying the $n\Delta\beta \cong \frac{\pi}{2} + 2m\pi, m = 0,1,2, ...$ condition. Therefore, there is an optimum value of $n$ for which the circuit can be truncated to obtain the best conversion loss. For example, for a NLTL based harmonic generator in [26] the number of 10 unit-cells are needed for a minimum of 9dB conversion loss operating at 3.5 GHz. The same concept can be applied for subharmonic generation which will be explained with more details in Sec. II. B. In general, for subharmonic generation circuitry, the number of stages defines both the resonant frequency, the resonator equilibrium condition and the input and output matching ports conditions. For instance, [38] proposed a 10 unit-cell configurations at higher frequency, 24 GHz, application, requiring high-impedance node for filtering out the subharmonics at the output node. This unilateral high impedance interface node is needed to isolate the output port from loading the resonator, which makes the method inherently non-bidirectional.

Therefore, in order to design a bidirectional harmonic and subharmonic circuitry, in this work the number of unit-cells and the output/input power matching units will be co-designed to simultaneously maintain the resonance conditions at the two bands and to maximize power transfer.

### B. Proposed Bidirectional Dual Band Non-Linear Ring Resonator

The standing wave resonator is the best candidate to sustain resonance conditions at both harmonic (input $f$, output $2f$) and subharmonic generation (input $2f$ and output $f$) in a closed-loop configuration, illustrated in Fig. 2 (d). In addition, the impedance matching design of the input and output ports can be integrated in the design of the resonator to create zeros or poles at the injection node depending on the frequency bands, $f$ or $2f$. In other words, the proposed NRR circuit, shown in Fig. 2 (d), should simultaneously maximize the power transfer while providing the resonance condition at the two bands.

To design the NRR circuity, one can start with the input impedance calculation of the circuit shown in Fig. 2 (d). It should be considered that the input voltage of the first cell is equal to the output voltage of the last cell while their currents flow with same amplitude but in the opposite directions. Based on transmission matrix technique, the input impedance at each node shown in Fig. 2 (d) can be calculated by deriving the relation of voltages and currents at every node as [32,46]

$$\begin{bmatrix} V_n \\ I_n \end{bmatrix} = \begin{bmatrix} A & B \\ C & D \end{bmatrix} \begin{bmatrix} V_{n+1} \\ I_{n+1} \end{bmatrix} \qquad (6.a)$$

And the A, B, C, and D values of the matrix can be found as

$$A = D = \cos \beta d - \pi f C_0 \sin \beta d \qquad (6.b)$$

$$B = j(\sin \beta d + \pi f C_0 \cos \beta d - \pi f C_0) \qquad (6.c)$$

$$C = j(\sin \beta d + \pi f C_0 \cos \beta d + \pi f C_0) \qquad (6.d)$$

where $C_0$ is the constant capacitance of each varactor, and $d$ is the transmission line half-arm length of each unit-cell and $\beta$ propagation constant that can be equated as

$$\cos \beta d = \cos k d - \pi f C_0 \sin k d \qquad (7)$$

where $k = 2\pi f \sqrt{\epsilon_0 \epsilon_r \mu}$ is the propagation constant of the simple transmission line, unloaded line, with dielectric constant, $\epsilon_r$, permeability $\mu$, and free space electric permittivity, $\epsilon_0$. Equation (7), plotted in Fig. 3, indicates that loading the transmission line with periodic $C_0$ capacitors increases $\beta d$, making the line as a slow-wave transmission line which is desired for low-frequency applications with miniaturized footprint such as IoT tags. Fig. 3 also specifies an approximate estimation of unit-cell electrical length at desired frequency, which will be further discussed in the resonator design expressed in Sec. II. C.

To estimate the input impedance, $Z_{in}$, at the injection node, $M$, in Fig. 2 (d), all the ABCD impedance matrix of the cascaded unit-cell can be multiplied to find the relation of $V_{in}/I_{in}$. The matrix-based impedance calculation requires the circuit operating in linear region, small signal mode, which can be an acceptable assumption for the onset of oscillation. Therefore, the I-V relation can be found by solving

$$\begin{bmatrix} V_1 \\ I_1 \end{bmatrix} = \begin{bmatrix} A & B \\ C & D \end{bmatrix}^N \begin{bmatrix} V_n \\ I_n \end{bmatrix} = \begin{bmatrix} A' & B' \\ C' & D' \end{bmatrix} \begin{bmatrix} V_1 \\ -I_1 \end{bmatrix} \qquad (8.a)$$

$$V_{in} = V_1 = V_N, \ I_{in} = 2I_1 = -2I_N, \ Z_{in} = \frac{V_{in}}{I_{in}}, \qquad (8.b)$$

To observe the resonance behavior of the circuit, one can plot the input impedance $Z_{in}$ versus frequency and as a function of the number of unit-cells $n$ of the NRR, illustrated in Fig. 2(d). As shown in Fig. 4, for a lossless circuit with pure imaginary impedance, at least three unit-cells are needed for the impedance to show resonance behaviors, corresponding to frequencies at which the impedance is either zero or infinite. Further increase in the number of unit-cells results in multiple resonant frequencies which might reduce the performance of the structure in terms of spurious signals and mixing products of the resonant frequencies. With three unit-cells, the input impedance has one zero around $f = 2.4\text{GHz}$, below the cut-off frequency $f_c$, Fig. 3, and one pole, infinite impedance, around second harmonics, $2f$. Around the zero-impedance frequency NRR resembles a series LC tank circuit, with the inductors and the capacitors cancel out their reactive impedances. While at the frequency pole near the second harmonic, $2f$, NRR has an infinite reactance resembling parallel LC circuit. The NRR circuit can resonate at these two bands which are not necessarily exact harmonics due to dispersion effect explained in (7) and shown as $\Delta \beta$ in Fig. 3. To make the generated resonance tones the exact harmonics of each other's, the power matching network with appropriate loading effect should be co-designed with the dual band resonator. The pumping power signals should also be symmetrically injected to the resonator through these power matched networks. Therefore, the trade-off between power transfer and loading effects on the resonator condition are needed to be taken care of. In the next section, the co-design of subharmonic divider and a novel input/output port for power matching filter will be discussed.

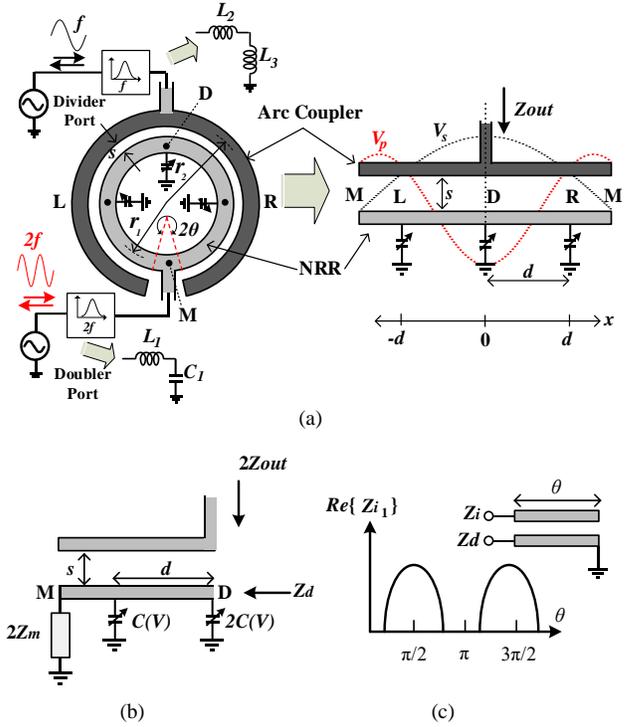

(a)

(b)                    (c)

Fig. 5: a) Proposed bidirectional nonlinear ring resonator with two excitation ports, and with the proposed center-fed arc shaped coupler with the standing waveforms along the NRR line (harmonic and subharmonic signals are shown by red and black colors respectively) b) Half-circuit analysis of the proposed arc coupled filter. c) Corresponding quarter wavelength band-pass filter characteristics.

### C. Co-design of bidirectional divider and novel arc-center coupler as a divider's power matching port

In order to have a sustainable subharmonic generation, a negative resistance needs to be generated at every diode in the unit-cell. The voltage waveform at a given node $n$ can be expressed as the superposition of subharmonic signal $v_{s,n}(\omega t)$ and the harmonic or pump signal, $v_{P,n}(2\omega t)$. By neglecting the possible time domain phase difference between harmonic and subharmonic signal, voltage at node $n$ for the proposed three unit-cell NRR, shown in Fig. 5 (a), can be expressed as

$$v_n(t) = v_{s,n}(t) + v_{s,n}(t)$$
$$= V_{s,n} \sin \omega t + V_{p,n} \sin 2\omega t \quad n = 1,2,3 \qquad (9)$$

Both $v_{s,n}(t)$ and $v_{p,n}(t)$ can be assumed as a couple identical waves propagating at the opposite directions. Therefore, the superposition of the two signals at node $n$ can be expanded as

$$
\begin{aligned}
\left| V_{s,n} \right| &= \left| V_{p,0} e^{\frac{j\beta_2 d}{2}} \left( e^{-j\beta_2 x} + e^{j\beta_2 x} \right) \right|_{x=\frac{(2n-1)d}{2}} \\
&= 2V_{p,0} \cos\left( \frac{(2n-1)}{2} \beta_2 d \right)
\end{aligned}
\tag{10}
$$

where $V_{p,0}$ is defined as the absolute amplitude of the pump signal, at $2f$, at node $M$. Equation (10) predicts that for a standing wave oscillation condition, an amplitude larger than the input amplitude is available at certain nodes to create a negative resistance through the diode. To calculate equivalent negative resistance $R_{e,n}$ of the diode at node $n$, one can place (10) in (4) and simplify the equation as (see. Appendix)

$$
R_{e,n} = -\frac{1}{\omega C_1 V_{p,0} \cos\left( \frac{(2n-1)}{2} \beta_2 d \right)}
\tag{11}
$$

Equation (11) shows that the negative resistance at any diode in the proposed NRR configuration is smaller than the resistance generated in the single diode-based divider circuitry or in a NLTL configuration with travelling wave propagation with equal pump voltage swing. The smaller negative resistance corresponds to a larger conductance, improving the input threshold power requirement significantly. This significant feature of the proposed circuitry is that it is functioning based on the standing-wave signal generation, creating larger voltage swing at certain nodes. The waveform of the harmonic, $v_P$, and the subharmonic signal, $v_S$, at every varactor node of the proposed NRR is illustrated in Fig. 5 (a). For the divider mode, the pump signal, $v_P$, has maximum swing at the divider port, node $D$, which is around $\sim 1.7V_{p,0}$ and around $\sim 1.3\ V_{p,0}$ at nodes $L$ and $R$, where two other diodes are located. The nodes at these locations receive the pump signal swing with amplitude larger than $V_{p,0}$, generating subharmonic tone, $v_S$, with maximum amplitude at the divider port, $D$, $V_{S,D}$, and amplitude of $0.86V_{S,D}$ at intermediate nodes, $L$ and $R$.

In order to extract each of these tones in a distributive way through the ring line and with minimum loading effect of the filter on the resonance condition (negative resistance generation), a novel center-fed arc shaped coupler is proposed as shown in Fig. 5(a). The proposed coupler has a symmetric layout for both harmonic and subharmonic with a maximum voltage swing occurring at its center, node $D$. Therefore, the circuit can be divided into two parallel identical half-circuits for the sake of analysis. The half-circuit schematic of the proposed coupler is shown in Fig. 5(b). For the subharmonic frequency generation, the impedance at node $M$, input pump port, is short circuit, as depicted in Fig. 4, transforming to open circuit, high impedance at the center node, $D$. It should be noted that this short to open impedance transformation occurs due to the half-circuit length of the ring which is equivalent to $\beta d$ of $\pi/2$, ($\beta d$ of each unit-cell is $\pi/3$, depicted in Fig. 3). On the other hand, by leaving the endpoint of upper coupled line as an open-stub, the configurations create a band-pass filter (BPF) with the impedance equations of $Z_i$ and $Z_d$ as [46]:

$$
Z_d = \frac{Z_0 Z_e}{Z_i},\ Z_i = \frac{\sqrt{Z_0 Z_e}\sqrt{(Z_0 - Z_e)^2 - (Z_0 + Z_e)^2 \cos^2\theta}}{(Z_0 + Z_e)\sin\theta}
\tag{12}
$$

where $Z_e$ and $Z_o$ are the even and odd mode impedances, respectively depending on the distance between the coupled line, $S$, the distributed capacitance and the characteristic impedance of each line, Fig. 5(b). For a simple transmission line supporting TEM waves, one can calculate these impedances related to these parameters. Nevertheless, for the proposed circuitry due to the asymmetric loading between outer line and inner line of the coupler, the equations will be more complicated and out of the scope of this paper. However, the band-pass filter (BPF) characteristic of the proposed arc coupler depends on the length of the half circuit line, $\theta$, as shown in Fig. 5(c) with a pass band centered at $\theta = \pi/2$ equivalent to the length of the proposed half-circuit ring. Therefore, the designated boundary condition of the nonlinear resonator requires short and open impedances at the interface nodes, making it a potential candidate to creates a band-pass filter based on the coupled line theory to extract the desired signal at the input/output ports simultaneously.

### D. Symmetrical and Co-design of the Doubler

Similar to the subharmonic generation in previous section, the same wave analysis can also be applied for the second harmonic signal generation due to the symmetry of the circuitry at node $D$ and $M$. In the doubler mode, the incoming signal is coupled to the NRR with the maximum coupling at node $D$ with $v_s(t) = V_{S,D}\sin\omega t$. This signal propagates in opposite directions along $x$. This mainly occurs due to the short circuit boundary condition at node $M$ set for subharmonic generation, creating open impedance at node $D$ and making the voltage signal bouncing back. This results in a standing wave signal with the normalized phasor that can be written as

$$
V_s(x) = \sin\left(\frac{2\pi x}{\lambda_1}\right) = \frac{A}{2j}\left( e^{j\beta_1 x} - e^{-j\beta_1 x} \right)
\tag{13}
$$

where $\lambda_1 = 2\pi/\beta_1$ is the wavelength of the subharmonic signal, and $e^{\pm j\beta_1 x}$ is the phasor term considering the effect of the two opposite propagating signals. Therefore, based on (5), the overall second harmonic, $V_P\sin 2\omega t$ will be the superposition of the second harmonics created by each of these two waves with opposite directions. By neglecting the loss factor $\alpha_2$ in (5), the total second harmonic can be written as

$$
V_{2f}(x) \cong k^{'}\left( e^{-j\beta_2 x} + e^{j\beta_2 x} \right) = 2k^{'}\cos\beta_2 x
\tag{14.a}
$$

$$
k^{'} = K_{C(v)}\frac{j\beta_2 nd}{\beta_2 - 2\beta_1}\sin\frac{(\beta_2 - 2\beta_1)nd}{2}
\tag{14.b}
$$

The derived equations indicate the waveform generated at each node is equal to the subharmonic mode, when the pumping signal at $2f$ propagating through the line in $x$ direction, expressed in (10). Therefore, by maximizing the input signal voltage swing at the input port of the doubler, node $D$ through the matching circuit, and with the existing boundary conditions of the divider mode, the similar waveform is expected to achieve at the second harmonic as shown in Fig. 5 (a). In addition, the number of the unit-cells can also be optimized to enhance the doubler voltage swing and conversion loss. However, due to the bi-directionality of the circuitry and avoiding other undesired resonances as shown in Fig. 4, the number of unit-cell is chosen to be $N$=3 for sub-6GHz application resulting in a pure spectrum with sustainable resonance boundary conditions.

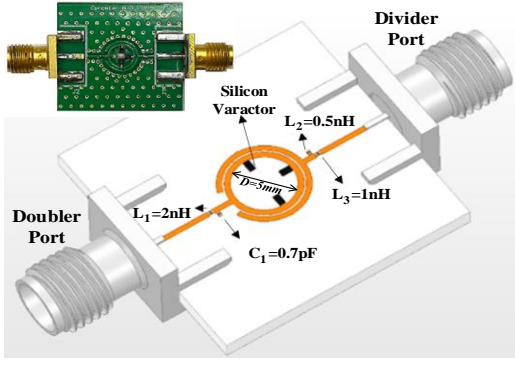

Fig. 6: PCB implementation of the proposed nonlinear ring resonator RO4003 substrate and with silicon abrupt junction nonlinear varactor.

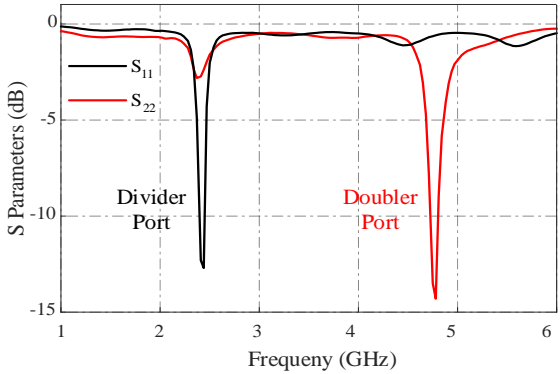

Fig.7: Measured S parameter of the bidirectional NRR at the two excitation ports, Divider Port ($S_{11}$) and Doubler Port ($S_{22}$) in Fig. 5.

## III. MEASUREMENT

The proposed circuitry is mounted on printed circuit board (PCB) with 8mil-RO4003 substrate and with $\epsilon_r = 3.5$ as shown in Fig. 6. The silicon abrupt junction varactor is employed due to its appropriate zero-bias capacitance $C_0$=2.67pF, resulting a low-profile resonator. With the unit-cell length of $d$=4mm the NRR will have a cut off frequency $f_c$=5.4 GHz, correspondent to the electrical length of $\pi/3$ at $f$=2.4 GHz, Fig. 3. This makes the three-unit-cell a half wavelength resonator with a inner ring diameter of 5 $mm$. The mm-size transmission line length makes the proposed system desirable for next generation miniaturized IoT sensor and tags. As it is equated in (7), the mm-scale size occurs due to enhancement of propagation constant, $\beta d$, in a periodic varactor loaded nonlinear line . In addition, the circuit is fully-passive running with zero bias voltage as the nonlinearity of $p$-$n$ junction varactor, $C_1$ coefficient in Fig. 2(a), is maximized at this biasing condition. This also eliminates the need for any energy harvesting circuitry and make the circuitry power-efficient candidate for low-power applications. The two ports of the circuit are also matched to 50Ω antenna ports through appropriate LC matching filter, Fig. 6. The doubler port is matched to the NRR with $L_1$=2nH and $C_1$=0.7pF to resonate at 4.8GHz while the divider port is connected to the center of the arc coupler line through $L_2$=0.5nH and $L_3$=1nH for maximum power transfer at 2.4GHz.

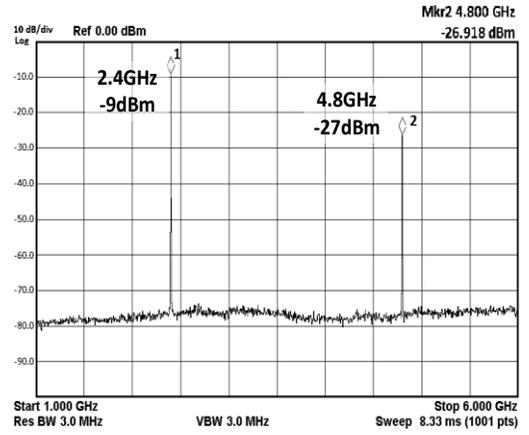

Fig. 8: Output power spectrum density of NRR circuit in the divider mode with 5dBm input applied to the port D at 4.8GHz.

The measured $S$-parameter of the two ports, $S_{11}$ at port $M$ and $S_{22}$ at port $D$, are plotted in Fig. 7. It indicates that the implemented circuitry provides two resonances at the two bands with input return loss better than 12 dB while suppressing the other band. The measured power spectrum density (PSD) of the circuit is also shown in Fig. 8, with 5 dBm input pump signal generating -9 dBm subharmonic power and about 32 dB intrinsic pump rejection at the divider port. This rejection is obtained by the filtering property of the proposed arc coupler explained in Fig. 5(c) without additional external filter.

The threshold power and the conversion loss of the divider are also shown in Fig. 9 with input pump power swept at 4.8 GHz. The measurement results show consistency with simulation results run in Advanced Design System (ADS). The required input threshold power, $P_{th}$, of the divider is around 1.5 dBm, which can be improved by tuning of the impedance matching filter lower than standard 50Ω. The simulation suggests that the required $P_{th}$ can be enhanced by 3dB if the output impedance of the arc coupler is translated to smaller impedance of 2Ω at the cost of output power reduction. The impedance transformation and matching optimization in divider mode result in conversion loss in the doubler mode due to bi-directionality. The frequency response of the circuitry in the divider mode is also shown in Fig. 10 under two different input pump power of 2dBm and 4dBm and compared by the simulation results. Both simulations and measurement confirm that increasing the pump power from 2 dBm to 4 dBm enhances the bandwidth from 20 MHz to 200 MHz, larger than 10 times per 2 dB power enhancement.

The measurement results for the doubler mode are illustrated in Fig. 11 and Fig. 12. By sweeping the input signal power at $f$=2.4GHz, the circuit generates second harmonic at 4.8 GHz with 18 dB conversion loss, while simulations predicted 10 dB conversion loss. The discrepancy between the simulation and measurement results of the doubler mode originates from mismatch and manufacturing tolerance effect of the transmission line. These affect the insertion loss and coupling factor of the coupler. The simulation results demonstrate that the 18 dB measured conversion loss is aligned with the simulation of the circuit with $s=0.4mm$ between the lines. By tuning the spacing, $S$, of the circuit, the coupling effect of the filter can be improved to achieve a conversion loss of 14 dB, shown in Fig. 11 and Fig.12. The results can be calibrated by trimming the distance $S$ in the structure or even by tuning the length of the arc-line. The measured frequency response of the doubler is shown in Fig. 12, with the 0 dBm input power for calibrated and un-calibrated version of PCBs. As it is illustrated, the output power of the doubler can increase by 4dB with x2 times enhancement of bandwidth by calibrating and trimming the transmission lines.

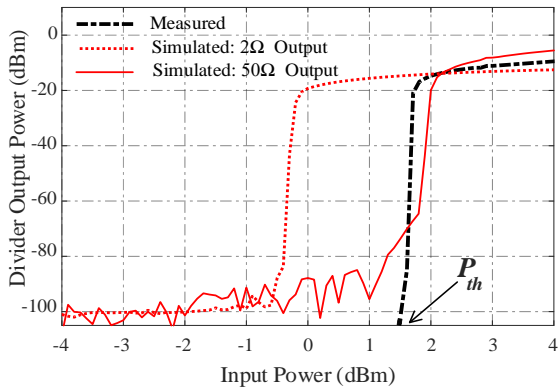

Fig.9: Measurement and simulation results for the output power in the divider mode at 2.4GHz output frequency vs input power swept at 4.8GHz input frequency.

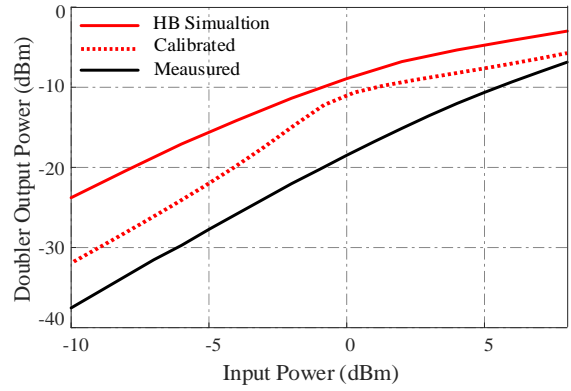

Fig. 11: Simulation and measurement results for second harmonic output power in the doubler mode vs input power at 2.4GHz input frequency.

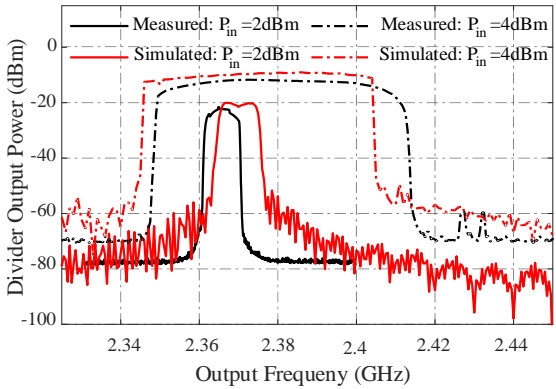

Fig. 10: Simulation and measurement results of the output power in the divider mode for 2dBm and 5dBm input powers.

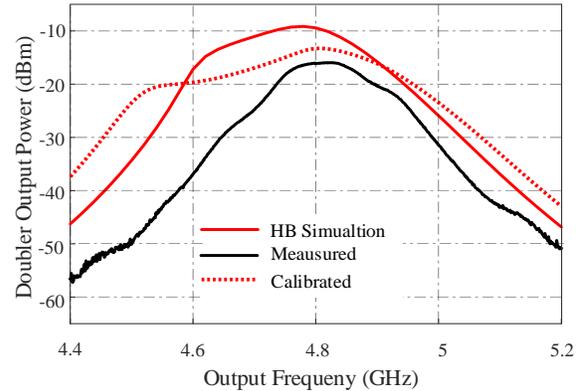

Fig. 12: Simulation and measurement results for second harmonic output power vs frequency for 0dBm input power.

The comparison of the proposed dual band circuitry is shown in Table I with other dividers [32-34] or doublers [23,26] employing nonlinear components for frequency generation. Firstly, it should be highlighted that this work is the first bidirectional circuit capable of generating and optimizing both harmonic, doubling, and subharmonic, dividing, with single circuitry and with only three unit-cells. The other periodic divider/doubler [26] and [34] requires at least 10 stage to generate the comparable results with conversion loss. The reduction of unit-cell size by using the repetitive cells in a ring configuration as well as zero-bias voltage requirement make the proposed circuitry a great candidate for miniaturized and low-power applications.

TABLE I COMPARISON BETWEEN THE PROPOSED CIRCUIT AND PASSIVE FREQUENCY DIVIDERS AND DOUBLERS

| | Type/ Implementation | $P_{th}$(dBm) | Conversion Loss @ 0dBm input | $P_{sat}$ (dBm) | $f_{out}$ (GHz) | Number of Stages | Bi-directionality | DC Bias (V) |
|---|---|---|---|---|---|---|---|---|
| This Work | Doubler/ | N/A | 14-18 | N/A | 4.8 | 3 | YES | 0 |
| | Divider (Lumped) | 1.5 | N/A | -9 | 2.4 | | | |
| [23] | Doubler (Lumped) | N/A | 14 | N/A | 2 | 1 | NO | 0 |
| [26] | Doubler (IC) | N/A | 10 | N/A | 6.8 | 10 | NO | 0 |
| [34] | Divider (IC) | ≃ 6 | N/A | -4.7 | 9.25-11.75 | 10 | NO | 0 |
| [32] | Divider (Lumped) | -10 | N/A | -20 | 0.1 | 1 | NO | 1.6 |
| [33] | Divider (Lumped) | -18 | N/A | -30 | 0.443 | 1 | NO | 0 |

## IV. CONCLUSION

This work presents a novel passive bidirectional frequency generation circuitry based on nonlinear ring resonator (NRR). The proposed circuit can translate the input frequency to both harmonics $(f \rightarrow 2f)$, and subharmonics, $(2f \rightarrow f)$ frequencies using minimum three number of nonlinear varactor unit-cell in a ring configuration. Due to the symmetry architecture of a ring resonator, the two resonance boundary conditions can be met at the two opposite excitation ports by creating short, zero, and infinite, pole, impedances. The proposed circuitry is implemented on PCB using RO4003 substrate and achieved 200MHz bandwidth in the divider mode with 10 dB and 16 dB averaged conversion loss in divider and doubler mode, respectively. The circuit consumes zero DC power with miniaturized 20 $mm^2$ layout.

## APPENDIX

The varactor capacitor is modulated by the pump voltage $v_p(t) = V_P \sin 2\omega t$. So, it can be approximated by the Taylor expansion around zero

$$C(v) = C_0 \left(1 + \frac{1}{V_J} V_P \sin 2\omega t\right)^{-m}$$

$$= \sum_{n=1}^{\infty} \frac{1}{n!} \frac{d^n C}{dv^n} = C_0 + C_1 v + C_2 v^2 + \cdots \quad (1)$$

Applying a small signal $v_s(t) = V_S \sin \omega t$ to varactor at the subharmonic frequency produces a current defined by

$$i_d = \frac{d}{dt} \left(C(v_p) \times v_s\right)$$

$$\simeq \frac{d}{dt} \left\{ \left(C_0 V_S + \frac{C_2 V_S V_P^2}{2}\right) \sin \omega t + \frac{C_1 V_S V_P}{2} (\cos \omega t - \cos 3\omega t) \right\} \quad (2)$$

Neglecting $\cos 3\omega t$ results in a summarized current as a function of $v_s(t)$ and $d/dt \, v_s(t)$

$$i_n \simeq \omega \left(C_0 + \frac{C_2 V_P^2}{2}\right) V_S \cos \omega t - \frac{\omega C_1 V_P}{2} V_S \sin \omega t$$
$$= C_e \frac{d}{dt} v_s(t) + \frac{1}{R_e} v_s(t) \quad (3)$$

Where the equivalent circuit of the varactor is capacitor $C_e$ in parallel with a negative resistor $R_e$.

$$C_e \simeq C_0 + \frac{C_2 V_P^2}{2}, R_e \simeq -\frac{2}{\omega C_1 V_P} \quad (4)$$

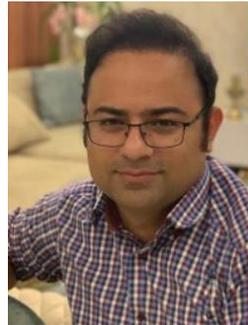

**Payman Pahlavan** received his B.Sc. degree in Electrical Engineering from K.N. Toosi University of Technology, Tehran, Iran, and his M.Sc. degree in Electrical Engineering from Sharif University of Technology, Tehran, Iran, in 2010 and 2013 respectively. He has more than seven years of industry experience in RF and Microwave research and development. He joined the University of Florida as a Ph.D. student. His research interests include RF and mm-wave integrated circuits, RF front-end circuit design, Internet of Things, and wireless communications.

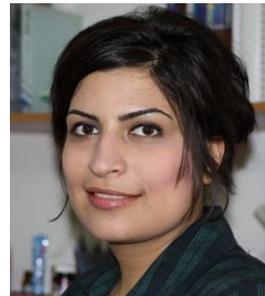

**Najme Ebrahimi** (S'09, M'17) received the B.S. degree (Highest Hons.) in electrical engineering from Shahid Beheshti University, Tehran, Iran, in 2009, the M.S. degree (Highest Hons.) in electrical engineering from the Amirkabir University of Technology, Tehran, in 2011, and the Ph.D. degree in electrical and computer engineering from the University of California at San Diego, La Jolla, CA, USA, in 2017. She was postdoctoral research fellow at the University of Michigan from 2017 to 2020. She is currently an Assistant Professor with the University of Florida, Gainesville, Florida, USA. Her research interests are on RF, millimeter-wave and THz Integrated circuits and systems, communication electronics, wireless communications and sensing, Internet of Things (IoT) connectivity and communications, physical layer security and sensing. Dr. Ebrahimi was a recipient of the Jacobs School of Engineering Fellowship at the University of California at San Diego, 2019 and 2020 EECS RisingStar, 2018-2020 IEEE Microwave Society Chapter Chair for Southeastern Michigan. Dr. Ebrahimi was a recipient of the 2021 Defense Advanced Research Projects Agency (DARPA) Young Faculty Award. She is as a member of the IMS2022 Technical Paper Review Committee (TPRC and is serving as a technical member of MTT-14 MICROWAVE AND MILLIMETER-WAVE INTEGRATED CIRCUITS.